# Cryogenic stabilization of molecular hydrogen in dense cubic ice

Tomasz Poręba[1,†], Leon Andriambariarijaona[2], Richard Gaal[1], Kazuki Komatsu[3], Gaston Garbarino[4], Thomas Hansen[5], Stanislav Savvin[5,6], Livia E. Bove[1,7,8,†]

1. Laboratory of Quantum Magnetism, Institute of Physics, École Polytechnique Fédérale de Lausanne, Lausanne CH-1015, Switzerland
2. Laboratoire pour l'Utilisation des Lasers Intenses (LULI), École Polytechnique, 91120 Palaiseau, France
3. Geochemical Research Center, Graduate School of Science, The University of Tokyo, Tokyo, Japan
4. European Synchrotron Radiation Facility, 71 avenue des Martyrs, 38000 Grenoble, France
5. Institut Laue-Langevin 71 avenue des Martyrs, CS 20156, 38042 Grenoble cedex 9, France
6. Instituto de Nanociencia y Materiales de Aragon, CSIC – Universidad de Zaragoza, Facultad de Ciencias C/ Pedro Cerbuna 12, Zaragoza, Spain
7. Institut de Minéralogie, de Physique des Matériaux et de Cosmochimie, Sorbonne Université, CNRS UMR 7590, MNHN, Paris, France
8. Dipartimento di Fisica, Sapienza Università di Roma, Roma 00185, Italy

† contact e-mail: tomasz.poreba@epfl.ch, livia.bove@sorbonne-universite.fr

**Hydrogen is widely regarded as a cornerstone of future low-carbon energy technologies, yet the lack of safe, efficient, and reversible solid-state storage materials remains a major barrier to its large-scale deployment. Although porous frameworks and metal hydrides have been extensively explored, far less is known about the ability of dense molecular solids to stabilize hydrogen at near-ambient pressure. Here we show that fully crystalline cubic ice, despite its non-porous nature, can retain molecular hydrogen as an interstitial guest following controlled decompression from a high-pressure hydrogen hydrate precursor. Using synchrotron X-ray diffraction, neutron diffraction, and Raman spectroscopy, we demonstrate that hydrogen is retained within the ice structure up to about 130 K, producing reproducible lattice expansion and distinct spectroscopic signatures. We further show that pure cubic ice can be partially refilled with hydrogen at 0.18 GPa and 130 K, while fully hydrogen-filled cubic structure can be preserved at the same pressure up to 90 K. The retained hydrogen content reaches several percent of the parent hydrate composition, corresponding to gravimetric and volumetric storage densities comparable to those of interstitial hydrogen in metals. These results reveal an unexpected ability of a dense hydrogen-bonded crystal structure to host molecular hydrogen without permanent porosity or chemical bonding, establishing cubic ice as a minimal model for hydrogen-lattice interactions. More broadly, our findings identify dense hydrogen-bonded solids as an unexplored class of materials for hydrogen storage physics, with implications extending from energy materials to planetary and astrophysical ice environments.**

Hydrogen's exceptionally high gravimetric energy density (119.9 MJ kg$^{-1}$), contrasts with its low volumetric energy density (9.8 kJ·dm$^{-3}$) under ambient conditions, severely limiting practical deployment as an energy carrier, and is driving the search for condensed-phase storage strategies based on physical confinement within crystalline solids.[1–3] To date, most approaches have focused on porous or chemically reactive hosts, including metal hydrides, zeolites, metal-organic frameworks, and templated carbons (Figure 1) which rely on either strong chemical bonding or confinement within permanent porosity.[4–8] While these materials can achieve significant hydrogen uptake, they are often limited by slow kinetics, incomplete reversibility, or the need for strong host-guest interactions that compromise cyclability. By contrast, dense molecular solids, in which hydrogen could be stabilized through weak, non-covalent interactions, remain comparatively unexplored. Hydrogen-bonded water networks offer a particularly intriguing platform in this context. Under pressure, water forms a variety of crystalline architectures capable of hosting small guest molecules, giving rise to hydrates.[9,10] In such materials, hydrogen is stabilized mainly by physical confinement within a hydrogen-bonded framework, and decomposition yields only water and molecular hydrogen - components that can be readily reintegrated into closed energy cycles.[11,12] However, under near-ambient conditions hydrogen solubility in both liquid water and hexagonal ice I$_h$, remains extremely limited (~1:30 H$_2$:H$_2$O even at 0.20 GPa).[13]

Application of pressure overcomes this limitation, enabling a rich family of hydrogen hydrate phases with distinct stoichiometries and host topologies (Figure 1 and Figure S1).[14–19] At moderate pressures (~0.25 GPa), and temperatures below 273 K hydrogen forms a cubic clathrate hydrate of structure II (sII) with an approximate 1:6 H$_2$:H$_2$O molar ratio. Further compression induces a sequence of structural transformations, culminating in the cubic C$_2$ structure (H$_2$:H$_2$O 1:1) above 3 GPa (inset Figure 1).[18,20] At still higher pressure (> 30 GPa), thermal activation of C$_2$ produces an even more hydrogen-rich cubic C$_3$ phase (2:1 H$_2$:H$_2$O).[17]

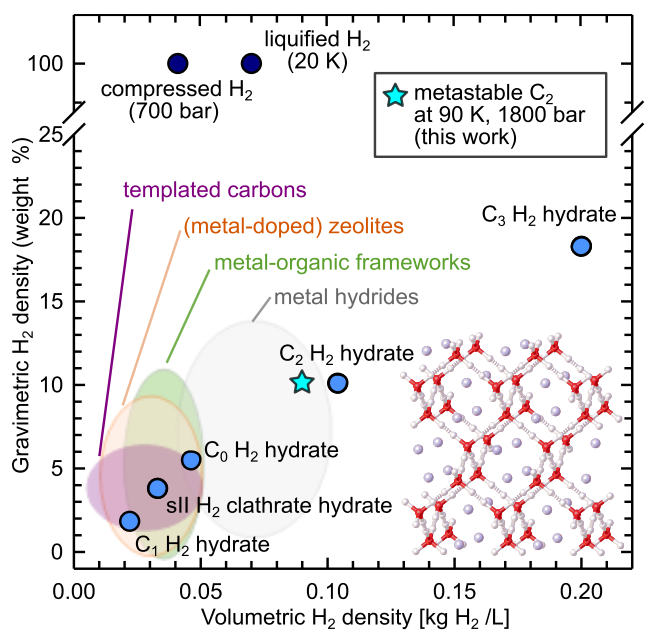

**Figure 1.** Gravimetric and volumetric hydrogen storage densities of known hydrogen hydrates at their lowest-pressure stability limits, compared with representative hydrogen-storage materials. Inset: crystal structure of $C_2$ hydrogen hydrate with guest $H_2$ represented as violet spheres.

The stability of these hydrogen-rich hydrates upon pressure release is a central issue for both fundamental understanding of weak hydrogen-water interactions and prospective storage concepts. While the sII clathrate can be preserved at ambient pressure under cryogenic conditions [21,22], the more hydrogen-rich $C_2$ phase exhibits reduced stability upon decompression, dissociating into cubic ice ($I_c$) and molecular hydrogen at sub-GPa pressures.[21] Notably, whereas cubic ice found in nature typically adopts a stack-disordered form ($I_{sd}$), composed of intergrown hexagonal and cubic layers[23], decomposition of hydrogen hydrates remains the only known route to produce stacking-fault-free ice $I_c$.[24–26] As a result, the recovered ice matrix retains a higher crystal symmetry, relative to hexagonal ice $I_h$, reflecting its high-pressure structural ancestry.

First neutron diffraction (ND) studies of decompressed $C_2$ reported no measurable hydrogen occupancy in the recovered $I_c$ at ambient pressure below 130 K, suggesting complete hydrogen release during decomposition.[24] However, other X-ray diffraction (XRD) and Raman spectroscopy studies suggested the formation of partially hydrogen-filled $I_c$, accompanied by lattice expansion and spectroscopic signatures consistent with retained molecular hydrogen.[24,27]

Despite these studies, the physical mechanism by which hydrogen can be stabilized within a dense, non-porous crystalline ice lattice - and the thermodynamic limits of such stabilization at ambient pressure - remain unresolved. It is unclear whether hydrogen retention in cubic ice reflects transient trapping associated with amorphous intermediates, or whether a genuine crystalline, hydrogen-bearing cubic ice phase can exist as a metastable solid under cryogenic conditions at ambient pressure.

Beyond terrestrial energy-storage considerations, this problem has broader implications for planetary and astrophysical ice chemistry. Cubic ice $I_c$ is expected to form under low-temperature, low-pressure conditions characteristic of cometary nuclei, icy satellites, and interstellar and circumstellar environments, where rapid freezing, irradiation, and pressure cycling promote the formation of metastable ice polymorphs.[28–30] This mechanism may be particularly relevant for icy bodies such as Enceladus - where molecular hydrogen has been directly detected in plume emissions - as well as for Europa, cometary nuclei, and interstellar grains, where irradiation-driven $H_2$ production coexists with the formation of cubic ice.[31] Under irradiation or energetic particle bombardment, water ice efficiently produces $H_2$, which is generally assumed to desorb rapidly or remain weakly adsorbed at grain surfaces.[28,32,33] Our results instead suggest that dense crystalline ice $I_c$ can act as a metastable host for molecular hydrogen, enabling its retention well above classical desorption temperatures and decoupling hydrogen storage from gas-phase thermodynamic equilibrium. In this context, cubic ice may function as a hidden hydrogen reservoir, capable of storing, transporting, and episodically releasing $H_2$ formed during serpentization processes occurring at depths.[34] This suggests that crystalline ice may act as a previously unrecognized reservoir for molecular hydrogen in icy planetary environments.

While hydrogen trapping has been extensively considered in amorphous ice and clathrate hydrates, the role of dense, non-porous crystalline ice phases as transient or metastable hydrogen hosts has received comparatively little attention.[35–39] Recognizing crystalline ice $I_c$ as an active participant in hydrogen retention therefore necessitates a reassessment of hydrogen budgets, volatile outgassing, and radiation-driven chemistry in cold planetary bodies and interstellar grain mantles, with implications spanning astrochemistry, planetary evolution, and the interpretation of hydrogen signatures in icy environments.[16,40,41]

In this work, we demonstrate that fully crystalline cubic ice derived from $C_2$ hydrogen hydrate can retain molecular hydrogen as an interstitial guest under cryogenic conditions (explored thermodynamic paths are presented in Figure S1). The hydrogen-stabilized $I_c$ phase exhibits reproducible lattice expansion and distinct spectroscopic signatures, persisting up to ~130 K at ambient pressure. The retained hydrogen concentration reaches several percent of the parent hydrate composition, corresponding to gravimetric and volumetric storage densities comparable to interstitial hydrogen in metallic systems. We further show that the parent $C_2$ phase can be partially preserved up to 90 K under 0.18 GPa.

# Results

## Phase evolution upon decompression

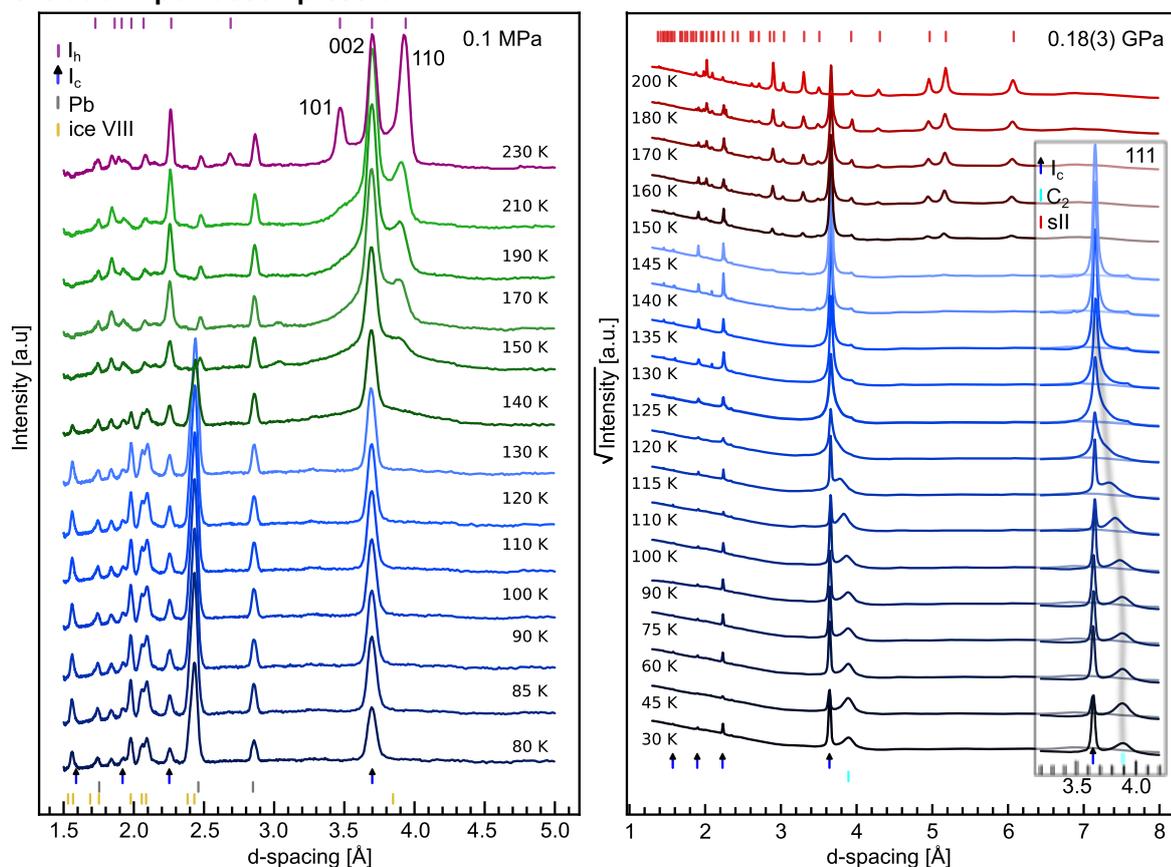

Hydrogen hydrates were synthesized through two independent routes (Experiment 1 and 2, respectively). In ND
**Figure 2.** (Left) Representative neutron diffractograms (ND, λ= 2.45 Å) of the $C_2$ sample decompressed to ambient pressure at 77K (Experiment 1) upon heating. (Right) X-ray diffractograms (XRD, λ= 0.40949 Å) of ice-free $C_2$ sample decompressed at 0.18(3) GPa at 30 K upon heating. Inset: Evolution of 111 Bragg reflection position for both $I_c$ and $C_2$. Grey line traces the unit-cell volume reduction through temperature-induced $H_2$ loss in $C_2$.

experiments, compression of deuterated sII hydrogen clathrate above approximately 3.2 GPa induced transformation into the filled-ice $C_2$ phase, accompanied by formation of excess ice VII (Experiment 1). Refinement of the guest occupancy yielded values close to unity (0.94(9)), confirming near-stoichiometric hydrogen filling consistent with previous studies.[16] A second synthetic route based on hydrolysis of $MgD_2$ (Experiment 2).followed by compression to 3.8 GPa produced the same $C_2$ hydrate, together with $Mg(OD)_2$ and residual unreacted ice VII (Figure S2).

Following rapid cooling to cryogenic temperatures and subsequent decompression to ambient pressure, both samples predominantly yield cubic ice $I_c$ with residual ice VIII (Figure 2, left panel). The persistence of cubic symmetry upon hydrogen evacuation reflects the topotactic relationship between $C_2$ and $I_c$, both of which crystallize in the *Fd-3m* space group with 8 formula units per unit cell. This structural continuity enables direct assessment of lattice changes associated with hydrogen removal without symmetry-breaking phase transitions (inset Figure 1).

Upon heating at ambient pressure, diffraction patterns evolve through a well-defined sequence (Figure 2 left). Below ~140 K, cubic ice reflections remain sharp. Around 140 K, diffuse features associated with amorphous ice emerge near the cubic 111 reflection, alongside the reduction in the Bragg intensities of residual ice VIII. With further heating, stack-disordered ice $I_{sd}$ develops, as evidenced by the appearance of broad shoulders adjacent to the cubic 111 peak that correspond to the 100 and 101 reflections of hexagonal ice $I_h$ (Figure 2 left). At ~230 K, $I_{sd}$ fully converts into hexagonal ice $I_h$, completing the transformation from hydrogen-bearing $I_c$ to the thermodynamically stable ambient-pressure phase.[42] In contrast to previous reports of hydrogen-filled cubic ice obtained via amorphous intermediates[42], these results show that cubic ice forms directly from an hydrogen-rich crystalline precursor below 140 K. The emergence of amorphous-like features is instead attributed to transformation of residual ice VIII rather than destabilization of the cubic lattice[43] Notably, this behaviour was mirrored in ND experiments from the samples produced by $MgD_2$ hydrolysis (Figure S3).

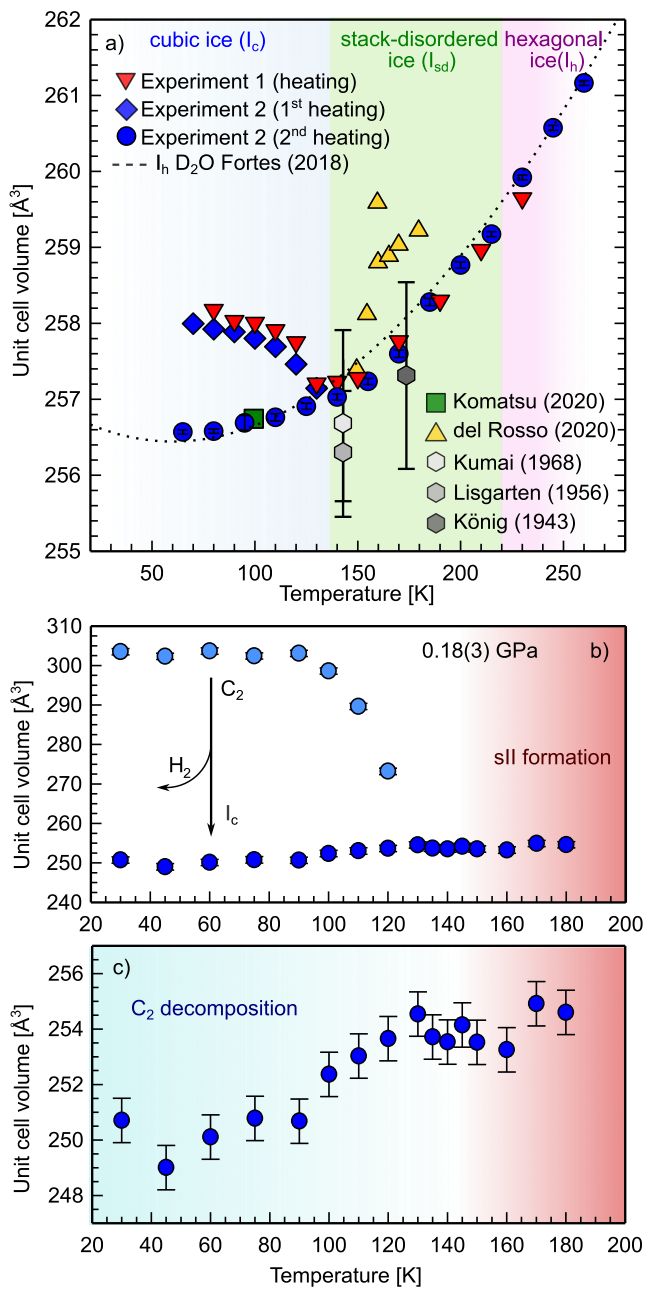

**Figure 3.** Unit cell volume evolution of cubic ice produced by decompression (to 0.1 MPa) of $C_2$ hydrogen hydrate at low temperature (red and blue symbols) determined by neutron diffraction (a). Nominal unit cell volume of $I_h$ is doubled for the visualization purpose. Thermal expansion in deuterated $I_h$ (dashed line)[68] is plotted along the experimental literature values for cubic ice unit cell volume.[26,69–71] For comparison, results of XRD experiments at 0.18(3) GPa in hydrogen-rich atmosphere show decomposition of $C_2$ (b) and simultaneous lattice expansion in $I_c$ due to hydrogen-refilling (c). Error bars, if not plotted, are smaller than the data point size

Ice-free $C_2$ hydrate was also obtained by compressing a hydrogen-loaded sample to 4.1 GPa in a diamond anvil cell (DAC), and its structure was confirmed by single-crystal X-ray diffraction (CCDC 2537162, Table 1 SI). Full decompression was not achieved due to mechanical constraints of the membrane DAC.

At the residual pressure of 0.18(3) GPa at 30 K, the $C_2$ phase decomposed partially into pure $I_c$ and hydrogen. A fraction of $C_2$ persists up to 90 K (Figure 2 right) under a hydrogen-rich atmosphere. Above this temperature, progressive hydrogen release leads to the complete conversion of the remaining $C_2$ into $I_c$ above 120 K, in line with the ND results.

Diffraction patterns of decompressed $C_2$ samples exhibit pronounced preferred orientation with only the low-angle 111 reflection clearly resolved, likely reflecting reduced crystallinity. Notably, the resulting $I_c$ does not develop stacking disorder at least up to 180 K, consistently with previous ND studies.[24] We further observe that $I_c$ reacts with excess hydrogen at 0.18(3) GPa already partially at 150 K, reforming sII clathrate. Consequently, only trace amounts of $I_c$ persist at 180 K (Figure 2, right panel). This also confirms that $I_{sd}$ observed in the ND experiments originates from ice VIII transformation above 130 K.[43]

**Hydrogen retention features**

Although refined guest $H_2$ occupancy values are statistically indistinguishable from zero, cubic ice obtained from decompressed $C_2$ exhibits a clear and reproducible unit-cell expansion relative to reference $I_h$ and $I_c$ phases (Figure 3a). At 80 K, the unit-cell volume is ~0.6% larger than that of $I_h$, when normalized to the same number of water molecules (consistently with the similar molecular volumes of $I_c$ and $I_h$). This excess volume cannot be explained by thermal effects or stacking disorder, as diffraction patterns show no diffuse features of $I_{sd}$. Instead, it indicates residual molecular hydrogen retained within the cubic ice framework at concentrations below the sensitivity of average-structure refinement.

Upon heating from 80 to ~130 K, the unit-cell volume decreases smoothly and converges toward that of hexagonal ice, consistent with progressive hydrogen release. Above 130 K the thermal expansion becomes indistinguishable from that of hydrogen-free ice $I_h$ (Figure 3a), indicating that guest-related contributions to lattice strain become negligible. Notably, thermal expansion of $C_2$ does not show any significant negative thermal effects (Figure S4).

To exclude intrinsic negative thermal expansion, we performed a temperature-cycling experiment (Experiment 2, Figure S3). $I_c$ was first heated to 130 K, where its volume matches $I_h$, then recooled to 50 K, and reheated. While the initial heating reproduces the anomalous contraction, this effect disappears entirely in the second cycle, with the lattice following the thermal expansion of $I_h$ (Figure 3a). These results establish that the low-temperature contraction arises from hydrogen release rather than intrinsic lattice effects. Therefore, fully crystalline $I_c$ retains molecular hydrogen below 130 K, independently of amorphous intermediates. The absence of volume differences between the heating cycles indicates that ice VIII does not retain $H_2$ whereas only $I_c$ accommodates measurable amounts of hydrogen (Figure S5).

Compression to 0.18(3) GPa enables partial retention of the $C_2$ phase up to 90 K (Figure 3b), demonstrating that $C_2$ can persist at low pressure, contrary to previous assumptions.[21]

Below 90 K, the unit-cell volume of $I_c$ remains essentially constant, apart from a single data point at 45 K. The abrupt volume decrease observed at this temperature, affecting both $I_c$ and $C_2$, is attributed to a pressure jump in the sample chamber associated with melting of excess hydrogen, which occurs near 46 K at 0.18 GPa.[44] This effect is attributed to a pressure jump associated with hydrogen melting near 46 K, within the uncertainty of ruby-fluorescence measurements.[45]

Above 90 K, a qualitatively different behavior emerges. As $C_2$ progressively decomposes, and releases $H_2$ into the sample chamber, the unit cell volume of $I_c$ increases linearly, reaching an expansion of 1.54% at 130 K (Figure 3c). Further heating in absence of $C_2$ does not produce any additional expansion. The observed volume increase (~3.8(7) Å$^3$) exceeds thermal expansion by more than an order of magnitude, providing compelling evidence for hydrogen re-uptake into the $I_c$ lattice.

Remarkably, this hydrogen-rich cubic ice phase remains stable up to 180 K at 0.18(3) GPa, with no detectable hydrogen loss. These results establish that $I_c$ can reversibly accommodate molecular hydrogen at moderate pressure and low temperature, overturning the assumption that hydrogen-filled $C_2$-derived structures cannot be preserved or re-formed outside high-pressure conditions.

**Spectroscopic evidence for hydrogen retention**

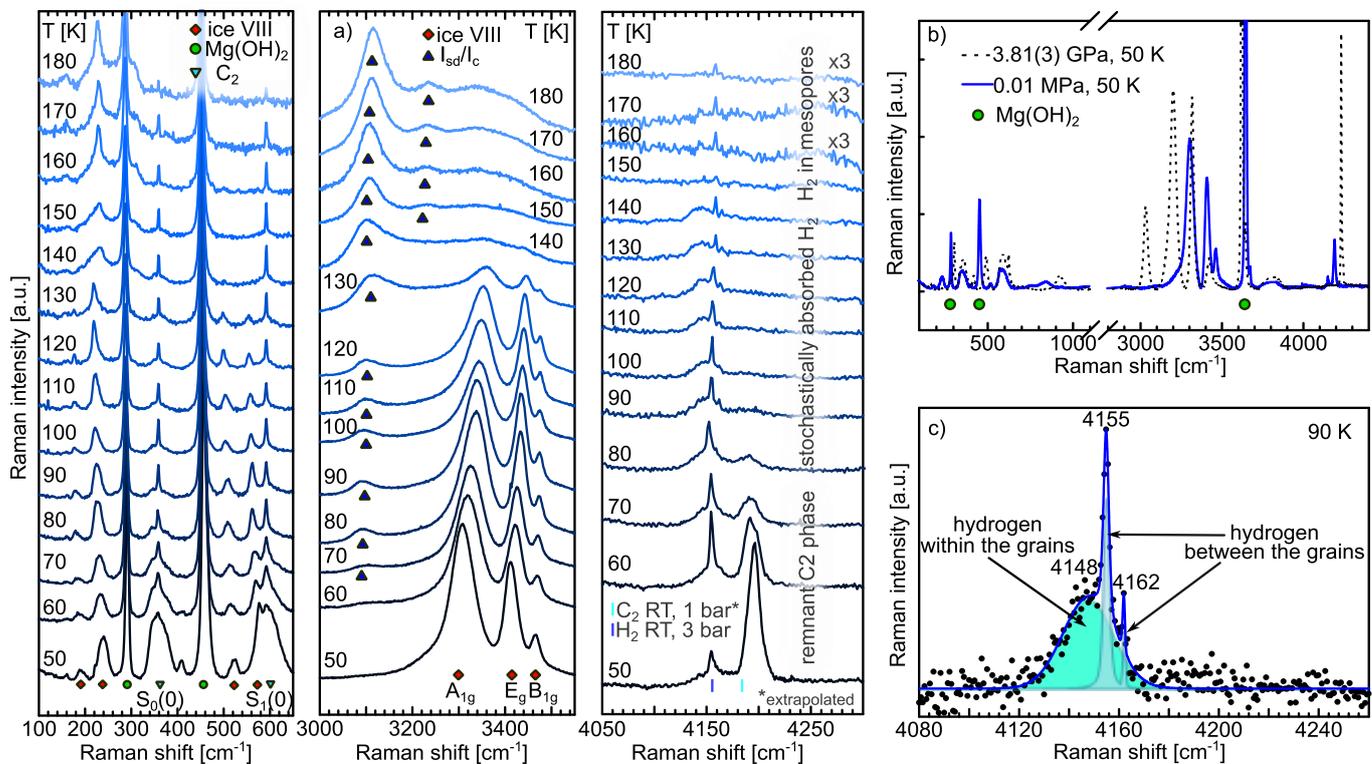

**Figure 4**. Raman spectra of decompressed $C_2$ hydrogen hydrate (a) with visible features from excess of ice VIII and Mg(OH)$_2$. Comparison of the Raman spectra before and after decompression at 50 K (b). Example of peak area integration in hydrogen-filled cubic ice at 90 K and ambient pressure; Gaussian function (teal) and two Lorenzian functions (blue) were used to integrate the signals from stochastically-adsorbed and intergrain molecular $H_2$, respectively (c).

Raman spectroscopy independently confirms the presence of molecular hydrogen retained within the cubic ice lattice. Spectra of decompressed $C_2$ samples at ambient pressure and cryogenic temperatures reveal weak but distinct $H_2$ roton and vibron features, superimposed on the lattice phonon and O-H stretching background of ice. The low-frequency region is dominated by translational modes of Mg(OH)$_2$, formed during MgH$_2$ decomposition (Figure 4a).

Between 50 and 90 K, the characteristic $S_0(0)$ and $S_1(0)$ hydrogen rotons are observed at 356 and 592 cm$^{-1}$, respectively (Figure 4a). These features are significantly broader than those of freely rotating $H_2$ in sII clathrate (42.8 vs 10.5 cm$^{-1}$ FWHM)[46], indicating occupation of multiple, locally distinct environments within the cubic ice framework. Above 90 K, the roton signals become weaker and sharper, reaching an average FWHM of 2.9 cm$^{-1}$, comparable to nearly free hydrogen.[46,47] Notably, an elevated spectral background extending ±25 cm$^{-1}$ around the roton position persists up to 150 K, coinciding with the complete transformation to hydrogen-free $I_c$ as determined by ND. The final disappearance of detectable $H_2$ roton features occurs only around 180 K, linked to trace amounts of weakly bound residual hydrogen within the $I_c$ matrix.

Characteristic ice VIII signal persists up to 130 K, before transforming to the one typical of $I_{sd}$[48], while the first signatures of $I_c$ appear near 70 K as a broad band centered at 3095 cm$^{-1}$ (Figure 4a).[49] This sequence mirrors the ND results, and confirms the absence of amorphous ice prior to $I_c$ formation.

The hydrogen vibron spectrum provides key evidence for $H_2$ incorporation into $I_c$. After decompression at 50 K, two vibron peaks are observed at 4197 and 4155 cm$^{-1}$, in contrast to the single strong peak at 4233 cm$^{-1}$ characteristic of stoichiometric $C_2$ at 3.81(3) GPa (Figure 4b). Vibron splitting and reduced intensity indicate partial hydrogen loss and occupation of two distinct local environments.

The higher-frequency vibron at (4197 cm$^{-1}$) is assigned to remnant $C_2$ domains, consistent with XRD observations and interpolated high-pressure Raman results.[17] The second peak, centered at 4155 cm$^{-1}$, coincides with the vibron of gaseous $H_2$ and persists up to at least 180 K (Figure 4a).[50]

Upon heating, the $C_2$-related vibron progressively weakens and vanishes, whereas the $H_2$-related band persists. Given the microlitre-scale sample volume, this signal cannot originate from hydrogen released into the cryostat. It instead reflects loosely bound $H_2$ retained within the sample. The initially sharp peak at 4155 cm$^{-1}$ develops a broader, Gaussian-like band centered near 4148 cm$^{-1}$ as $C_2$ fully transforms into hydrogen-filled $I_c$ (Figure 4c). The sharp "free-hydrogen-like" vibron is attributed to $H_2$ trapped in intergranular regions, whereas the broader Gaussian component reflects hydrogen distributed within $I_c$ grains (Figure 4c). When heated, hydrogen is progressively desorbed first from the crystallite interior and subsequently from intergranular regions, where it may remain mechanically trapped. Similar Raman signatures have been reported for hydrogen-filled amorphous silica microspheres during low-temperature hydrogen desorption, where the similar broad vibron component was attributed to stochastically physisorbed $H_2$.[51]

Overall, the Raman data identify three distinct processes upon heating: (i) decomposition of remnant, partially filled $C_2$ below ~90 K; (ii) hydrogen desorption from stochastically filled $I_c$ up to ~140 K; and (iii) hydrogen release from intergranular mesopores at higher temperatures. This sequence closely mirrors the ND results, where the unit-cell volume contraction associated with hydrogen desorption is complete by ~140 K.

**Hydrogen filling fraction**

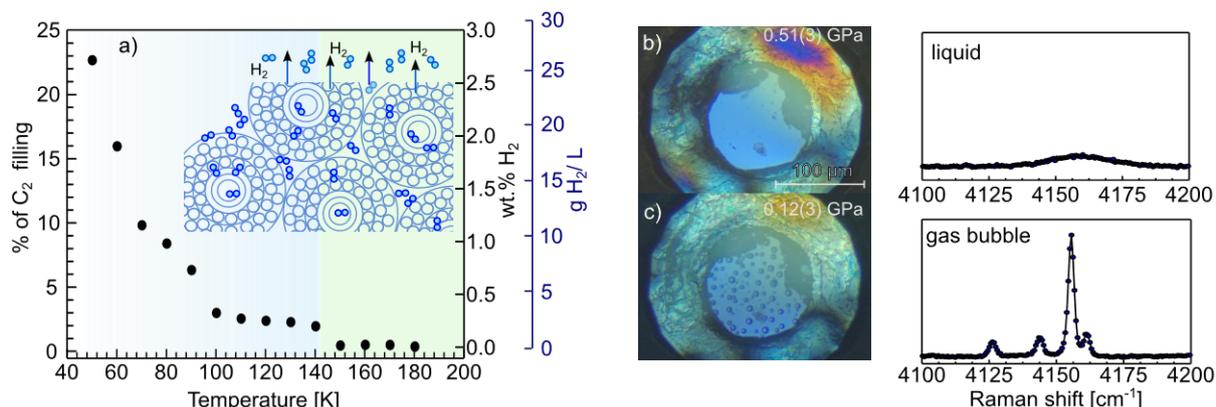

**Figure 5.** Hydrogen filling fraction in decompressed $C_2$ determined by Raman signal integration of $H_2$ vibron, along with corresponding gravimetric and volumetric $H_2$ densities (a). Inset: schematic representation of guest $H_2$ molecules distribution within and between the $I_c$ grains, and $H_2$ loss upon heating. Background color corresponds to stability regions of pure $I_c$ and $I_{sd}$ determined from neutron diffraction. Microphotographs of DAC chamber filled with recovered hydrogen-filled $I_c$ recompressed at 0.51(3) GPa and 300 K (b), and the same sample subsequently decompressed to 0.12(3) GPa (c). Corresponding Raman spectra collected in the liquid (0.51(3) GPa) and evolved $H_2$ bubble (0.12(3) GPa), respectively, are presented in right-hand panel.

Quantitative analysis of the integrated vibron intensities reveals a gradual decrease in hydrogen signal intensity upon heating, closely paralleling the contraction of the cubic unit cell. This correlation demonstrates a direct coupling between hydrogen content and lattice strain, showing that hydrogen occupies interstitial sites within the ice framework rather than segregating into macroscopic inclusions. A fraction of hydrogen, however, remains trapped in mesoporous regions generated during $C_2$ decomposition upon decompression (inset Figure 5a).

Near-zero $H_2$ occupancies from ND reflect the loss of long-range order rather than absence of $H_2$. The presence of interstitial hydrogen can instead be inferred from the observed lattice expansion. Comparison of molecular volumes of $C_2$ and hydrogen-bearing $I_c$ yields an estimated lattice expansion of ~6.5 Å$^3$ per $H_2$ molecule, comparable to interstitial hydrogen in *fcc* metals (~5.8 Å$^3$).[52,53] Based on this $H_2$ molecular volume estimate, the hydrogen filling fractions in $I_c$ at 70 K correspond to 3.06% and 2.48% of the $C_2$ stoichiometry ($H_2 \cdot H_2O$ 1:1) for Experiments 1 and 2, respectively.

On the other hand, analysis of the integrated Raman vibron intensities yields higher apparent hydrogen contents of 9.9% at 70 K, increasing to 22.7% for samples decompressed at 50 K (Figure 5). The higher apparent hydrogen content from Raman experiments likely reflects partial thermal exposure during sample transfer. Raman spectroscopy experiments indeed show that most of the remnant $C_2$ decomposes near 90 K, reducing the hydrogen filling of $I_c$ to 3.04%, in excellent agreement with the filling fractions inferred from diffraction. Altogether, Raman spectroscopy and diffraction experiments unambiguously establish $H_2$ as an interstitial guest in $I_c$, with its presence manifested through lattice strain rather than long-range crystallographic order.

Pressure stabilizes the $C_2$ down to 90 K at 0.18(3) GPa but also enables re-incorporation of $H_2$ into the $I_c$ lattice. Quantifying the hydrogen filling is challenging due to the combined effects of pressure uncertainties inherent to DAC measurements and the low bulk modulus of ice, which together lead to large uncertainties in unit-cell volumes (up to ±0.8 Å$^3$). Nevertheless, relative volume differences between $C_2$ and $I_c$ provide a robust basis for estimating hydrogen uptake. From these data, we infer a maximum refilling level of 6.6 ± 1.5% relative to fully occupied $C_2$ between 130 and 180 K at 0.18(3) GPa. Effervescence of $H_2$ is directly observed in $C_2$ samples decompressed to ambient pressure at 77 K and subsequently heated up to 300 K (Figure 5b). $H_2$ gas bubbles formation observed upon decompression to 0.12(3) GPa is further confirmed by micro-Raman spectroscopy (Figure 5c).

The extrapolated (using reported isothermal compressibility[54] of $I_h$) average unit-cell volume of 258.7(8) Å$^3$ in this temperature range at ambient pressure closely matches that of ice $I_c$ previously obtained from degassed and annealed $C_0$ hydrogen hydrate (Figure 3a).[26] This suggests that previously reported "pristine" $I_c$ may have contained residual hydrogen.

More generally, hydrogen incorporation is governed by hydrogen fugacity rather than pressure alone. Elevated pressure increases the chemical potential of external $H_2$ relative to that of lattice-bound hydrogen, making refilling of $I_c$ thermodynamically favorable (Figure S6). Consistent with this picture, the transformation of $C_2$ ($H_2 \cdot H_2O$) to the more hydrogen-rich $C_3$ phase ($2H_2 \cdot H_2O$) at very high pressure (~40 GPa) has been shown to occur only in the presence of fluid hydrogen [17] In the absence of an external hydrogen, $C_2$ instead decomposes into ice VII and molecular hydrogen.[55]

In conclusion, hydrogen-filled cubic ice does not compete with conventional storage materials in capacity but defines a fundamentally distinct paradigm: reversible hydrogen storage in a dense, non-porous molecular crystal at ambient pressure, without covalent bonding, extreme pressures, or structural collapse. With gravimetric (~1 wt.% at 77 K) and volumetric densities comparable to interstitial hydrogen in metals, $I_c$ provides a clean model system for probing weak hydrogen-lattice interactions and the limits of physisorption in crystalline solids.

Beyond materials science, these findings have implications for planetary and astrophysical ices. Cubic ice can transiently stabilize molecular hydrogen under low temperature non-equilibrium, low-temperature conditions relevant to cometary nuclei, icy satellites, and interstellar grains. This mechanism - distinct from clathrate encapsulation or amorphous trapping - defines a new pathway for hydrogen retention and delayed release in cold planetary environments. Metastable crystalline ice phases thus emerge not as passive byproducts, but as active reservoirs in hydrogen–water chemistry under extreme conditions.

## Methods

### X-ray diffraction

In-situ synthesis of hydrogen hydrate and subsequent diffraction experiments were performed using a membrane-driven diamond anvil cell (DAC) equipped with diamonds with 350 μm-diameter culets. A sample chamber consisted of a 180 μm-diameter hole drilled in a Re gasket pre-indented to 70 μm. To minimize hydrogen diffusion into the gasket, the rim of the sample chamber was protected by metallic gold ring. A ruby was placed close to the gasket was used as a pressure-calibrant.[56] A drop of Millipore (18.2 MΩ·cm) water was placed inside the gasket hole was left to evaporate till circa 70% of the sample chamber was filled with air bubble. Then, the DAC cover was firmly closed to stop further evaporation. The DAC was opened pneumatically inside the gas-loading system, and the air bubble was exchanged with 1400 bar of 99.99+% $H_2$. Synthesis of $C_2$ was performed by compressing the mixture to 4.10(3) GPa. The identity of the sample was confirmed by single-crystal X-ray diffraction. The DAC was placed in an evacuated (~10$^{-6}$ Torr) He cryostat, and the pressure inside the DAC membrane was controlled using PACE5000 pressure controller. The powder diffraction experiments were conducted at the ID15B beamline at the European Synchrotron Radiation Facility using a monochromatic X-ray beam with a wavelength of 0.4100 Å and an FWHM of approximately 2.5 × 2.5 μm size.[57] Single-crystal data was collected using a wavelength of 0.3738 Å. The wavelength was calibrated using a CeO$_2$ powder standard. Single-crystal data were collected using an Eiger 9M CdTe detector either through ±35° ω-scans and a step size of 0.5° around one rotation axis (single-crystal data) or through continuous ±10° ω-scans (powder diffraction data). After transformation into the Esperanto format, the CrysAlisPro program suite was used for the indexing and data reduction for single-crystal data.[58] The structures were refined with Shelxl, as incorporated in OLEX2.[59,60] Powder diffraction data was integrated and processed in Dioptas software.[61] Le Bail refinements of resulting diffractograms were performed in Topas v7 software.[62] CCDC 2537162 contains the supplementary crystallographic data for this paper. The data can be obtained free of charge from The Cambridge Crystallographic Data Centre via www.ccdc.cam.ac.uk/structures.

## Raman spectroscopy

Preparation of $C_2$ hydrogen hydrate was performed by placing 1:4 w/w mixture of $MgH_2$ (synthesized according to procedure described in reference 21) and $H_2O$ (Milli-Q, 18.2 MΩ·cm) inside a membrane DAC equipped with IIas class diamonds, prepared in a same way as in X-ray diffraction measurements (see above). Additionally, to ensure the full opening of the DAC upon decompression, stacks of Belleville conical gaskets were placed on the pillars connecting the top and the bottom DAC plates, in a way that they ensured 0.2 mm gap when not pressed. The DAC was pressurized to 0.3 GPa and placed in a laboratory oven at 85°C for 2 hours. After cooling down to room temperature, the reaction progress was checked using Raman spectroscopy by monitoring the integrated intensity of $H_2$ vibron around 4160 cm$^{-1}$. $C_2$ was produced by isothermal compression to 3.81(3) GPa at 300 K. Raman spectra were collected using argon laser (120 mW output power) tuned to 514.32 nm and a Peltier-cooled HR460 spectrograph with 1500 gr/mm grating. The acquisition time was typically about a minute. In the Raman measurements, pressure was determined by the shift of the R1 ruby fluorescence line. As a reference, one ruby sphere was placed on the outer surface of the diamond, using heat-conducting paste. Helium pressure inside the DAC membrane was continuously controlled using PACE5000 pressure controller. An evacuated (around 10$^{-5}$ Torr) helium cryostat was used for low-temperature measurements. The DAC was fixed in the cooling block using bronze screws and minimal amount of vacuum-proofed thermal paste.

## Neutron diffraction

Experiment 1

Preparations of sII hydrate were performed at the Helmholtz-Centre Berlin (Germany) by exposing ice crystallites at 244 K to $D_2$ gas at 0.28 GPa for about 6 hours. Ice spheres with typical diameters of 10-30 μm were formed by spraying $D_2O$ into liquid $N_2$ under inert atmosphere ($N_2$).
The deuterated sII sample of about 30 mm$^3$ was loaded cryogenically (80 K) into a toroidal TiZr gasket along with a Pb pressure marker inside the locking clamp device of the Paris-Edinburgh press.[63] Subsequently, the clamp was tightened to exert pressure on the sample (~0.2 GPa) and then transferred into the press. Pressure was determined from equation of state of Pb[64]. The pressure was ramped up to 4.1 GPa and then the press was cooled rapidly by plunging into liquid $N_2$ bath. The system was decompressed, and the sample inside the gasket was transferred into a vanadium can submerged in liquid $N_2$. Then the can was transferred into an evacuated standard orange cryostat.[65] The neutron diffraction data was collected at the diffractometer XTREMED at the Institut Laue-Langevin (Grenoble, France), using a wavelength of 2.45 Å (experiment DOI: 10.5291/ILL-DATA.5-22-833).[66] Le Bail and Rietveld refinements of the collected diffractograms was performed in Topas v7 software.[62]

Experiment 2

Preparation of C2 hydrogen hydrate was performed by placing 1:4 w/w mixture of $MgD_2$ (synthesized according to procedure described in reference 21) and $D_2O$ (99.9 atom % D, Sigma-Aldrich) inside a toroidal TiZr gasket placed inside liquid $N_2$ bath. The gasket was then transferred into the locking clamp device of the Paris-Edinburgh press. Subsequently, the clamp was tightened to exert pressure on the sample (~0.3 GPa) and then transferred into the press. The center of the press where the gasket was placed was heated to 150°C with a heat gun for 2 hours to initiate the following nominal reaction:

$$MgD_2 + 3D_2O \rightarrow 2D_2 + D_2O + Mg(OD)_2$$

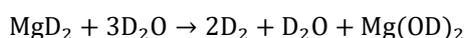

After cooling down to room temperature, the clamp module was mounted in Paris-Edinburgh press and compressed to 3.8 GPa. The following steps after cooling down in liquid nitrogen bath were the same as in Experiment 1. The ND data were collected at the high-intensity diffractometer D20 at the Institut Laue-Langevin (Grenoble, France), using a wavelength of 1.54 Å.[67]

## Acknowledgements

We are grateful to Dr. Maria Rescigno and Alasdair Nicholls for assisting during neutron scattering experiments, and Dr. Sophie Espert and Loan Renaud for assisting during synchrotron XRD experiments. We acknowledge the European Synchrotron Radiation Facility for provision of synchrotron radiation facilities under proposal number HC-5914 and we would like to thank S. Gallego Parra for assistance and support in using beamline ID15b. We acknowledge Institut Laue-Langevin for provision of neutron radiation facilities: D20 and XtremeD (proposals 525295 and 522833, respectively). We acknowledge U. Ranieri, A. Falenty, W. Kuhs, and L. Ulivi for sample preparation. L.E.B. acknowledge the financial support by the European Union–NextGenerationEU (PRIN N. 2022NRBLPT) and the ANR-23-CE30-0034 EXOTIC-ICE. T.P and L.E.B. acknowledge the financial support by the Swiss National Funds (grant number: 212889). K.K. acknowledge the financial support by the JSPS KAKENHI (No. 21K18154 and 25K01724), the Mitsubishi Foundation (No. 202310023),

**Data Availability**

The data supporting the findings of this study are available within the Article and its Supplementary Information. Other information related to this study is available from the corresponding authors upon reasonable request.